\documentclass[sigconf, screen, nonacm]{acmart}

\AtBeginDocument{%
  }

\setcopyright{acmlicensed}
\copyrightyear{2018}
\acmYear{2018}
\acmDOI{XXXXXXX.XXXXXXX}
\acmConference[AVI 2026]{International Conference on Advanced Visual Interfaces}{June 8--12, 2026}{Venice, Italy}
\acmISBN{978-1-4503-XXXX-X/2018/06}

\usepackage{multirow}
\usepackage{xspace}
\usepackage{stfloats}
\newcommand{\HUMA}{HUM\textsubscript{\emph{a}}\xspace}
\newcommand{\HUMR}{HUM\textsubscript{\emph{r}}\xspace}
\newcommand{\GPTA}{GPT\textsubscript{\emph{a}}\xspace}
\newcommand{\GPTR}{GPT\textsubscript{\emph{r}}\xspace}
\usepackage{subcaption}

\begin{document}

\title{The Effect of Idea Elaboration on the Automatic Assessment of Idea Originality}

\author{Umberto Domanti}
\affiliation{%
  \institution{Free University of Bozen-Bolzano}
  \city{Bolzano}
  \country{Italy}}
\email{udomanti@unibz.it}

\author{Moritz Mock}
\affiliation{%
  \institution{Free University of Bozen-Bolzano}
  \city{Bolzano}
  \country{Italy}}
  \email{momock@unibz.it}

\author{Sergio Agnoli}
\affiliation{%
  \institution{University of Trieste}
  \city{Trieste}
  \country{Italy}}
  \affiliation{%
    \institution{Marconi Institute for Creativity}
    \city{Bologna}
    \country{Italy}}
\email{sergio.agnoli@units.it}

\author{Antonella De Angeli}
\affiliation{%
  \institution{Free University of Bozen-Bolzano}
  \city{Bolzano}
  \country{Italy}}
  \email{antonella.deangeli@unibz.it}

\renewcommand{\shortauthors}{Domanti et al.}

\begin{abstract}
Automatic systems are increasingly used to assess the originality of responses in creative tasks. They offer a potential solution to key limitations of human assessment (cost,  fatigue, and subjectivity), but there is preliminary evidence of a self-preference bias. Accordingly, automatic systems tend to prefer outcomes that are more closely related to their style, rather than to the human one. In this paper, we investigated how Large Language Models (LLMs) align with human raters in assessing the originality of responses in a divergent thinking task. We analysed 4,813 responses to the Alternate Uses Task produced by higher and lower creative humans and ChatGPT-4o. Human raters were two university students who underwent intensive training. Machine raters were two specialised systems fine-tuned on AUT responses and corresponding human ratings (OCSAI and CLAUS) and ChatGPT-4o, which was prompted with the same instructions as human raters. Results confirmed the presence of a self-preference bias in LLMs. Automatic systems tended to privilege artificial responses. However, this self-preference bias disappeared when the analyses controlled for the idea elaboration. We discuss theoretical and methodological implications of these findings by highlighting future directions for research on creativity assessment.
\end{abstract}




\begin{teaserfigure}
    \centering
    \includegraphics[width=\linewidth]{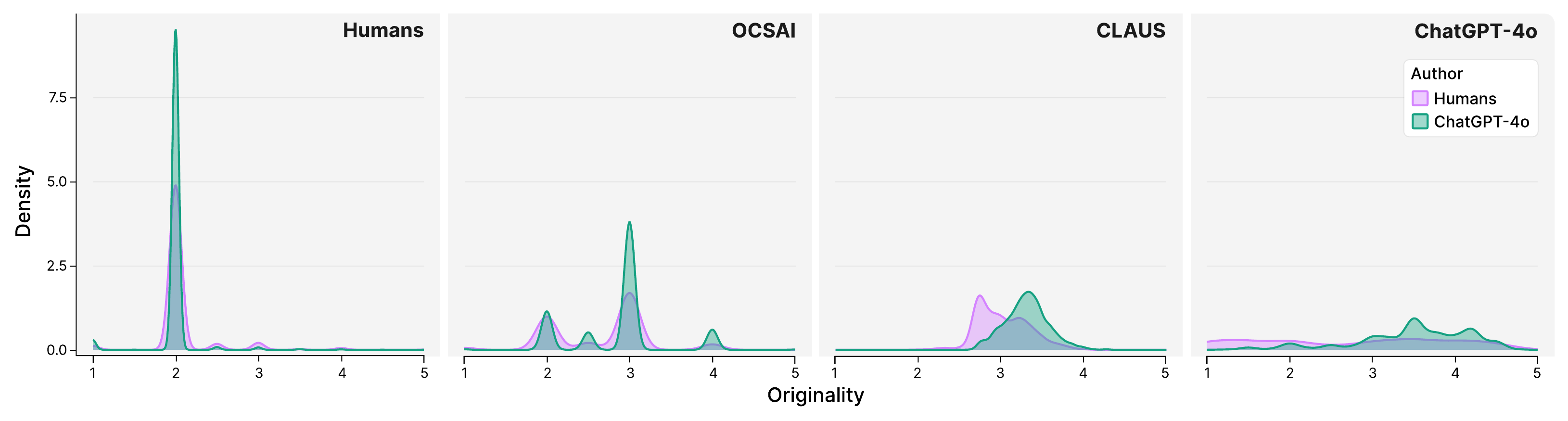}
    \Description{Figure 1 shows kernel density estimates of Alternate Uses Task originality scores of Human participants and ChatGPT-4o, shown separately for four raters: Humans, OCSAI, CLAUS, and ChatGPT-4o. Each panel displays two density curves (Humans and ChatGPT-4o). The X-axis shows the originality scale and it ranges from 1 to 5. The Y-axis represents the density. It ranges from 0 to 7.5. For each rater two density curves are shown. For human raters, the curve of human and ChatGPT-4o are concentrated around a single scale point (2.0). For OCSAI, the two curves are characterised by a central tendency around 3.0. For CLAUS, the two curves are narrower and more symmetric. For ChatGPT-4o as a rater, the two curves are widely distributed on the scale.}    
    \caption{Kernel density estimates of initial response originality scores across raters (Humans, OCSAI, CLAUS, ChatGPT-4o)}
    \label{fig:full_density}
\end{teaserfigure}


\maketitle
\section{Introduction}
Researchers in the field of creativity have recently begun to explore the various ways in which Artificial Intelligence (AI) can shape human behaviour. In their manifesto, Vinchon et al. \cite{vinchon_manifesto_2023} envision four possible scenarios: 1. “Co-cre-AI-tion”, the output would not be possible by humans or AI alone; 2. “Organic”, creation made by humans for humans; 3. “Plagiarism 3.0”, users will draw on AI production without citing the source; 4. “Shut down”, users will become less motivated to start a creative action \cite{vinchon_manifesto_2023}. Creativity was originally defined as the human ability to produce original and effective outcomes or products \cite{runco_divergent_2012}. Generative AI (GenAI) has challenged this definition, requiring the addition of intentionality and authenticity to differentiate between human and \emph{artificial creativity} \cite{RUNCO_2023, Runco_2025}. 

The study of creativity addresses two main processes:  generation and assessment. In this paper, we focus on the assessment of idea originality in divergent thinking tasks. A prototypical measure of this ability is the Alternate Uses Task \cite[AUT;][]{GUILFORD}, in which participants are asked to generate as many alternative uses as possible for common objects (e.g., brick, pants, tyre) within a limited amount of time.  
Traditionally, responses have been assessed by human experts \cite{silvia_assessing_2008}; however, it has recently been suggested that Large Language Models (LLMs) could offer a solution to the limitations of human originality assessment. Yet, important concerns remain regarding the potential presence of biases, which might lead to inaccurate assessments \cite{patterson_cap_2025, Tang_2025}. One of these is the self-preference bias, in which AI-generated responses are preferred over the human ones \cite{Tang_2025, self_llm_2024}. In addition, even though LLMs have shown performances comparable to those of human raters in assessing originality, they still fail to replicate central dimensions like human experience and emotional depth \cite{beaty_chapter_2026}. 
The research question addressed in this paper ~is:
\begin{description}
    \item[RQ]  \emph{How do human and artificial raters differ in the assessment of idea originality?}
\end{description}
To answer it, we investigated how two fine-tuned systems (OCSAI, CLAUS) and ChatGPT-4o align with human raters in scoring the originality of AUT ideas generated by ChatGPT-4o and human authors. The paper has the following organisation. Section \ref{related_work} presents the related work about human and automatic assessment of originality. Section \ref{sec_method} presents the method of the study, and Section \ref{sec_results} reports the results. Section \ref{sec_discussion} discusses their theoretical and methodological implications, limitations,  and highlights future directions. Lastly, Section \ref{sec_conclusion} concludes the paper and links the main findings to the literature.

\section{Related Work} \label{related_work}
Divergent thinking is the ability to generate multiple alternative solutions to a problem \cite{acar_divergent_2019}. Four indices are typically measured in its assessment \cite{acar_latency_2019}. \emph{Fluency} is operationalised by counting the number of ideas provided by an individual. \emph{Flexibility} refers both to the number of different categories of ideas and the number of switches between conceptual fields. 
\emph{Elaboration} reflects the amount of detail provided in the generated idea. Finally, \emph{originality} represents the most frequently analysed aspect of divergent thinking, and is usually assessed by individual trained raters \cite{silvia_assessing_2008}. The focus of this paper is on originality and elaboration.

\subsection{Human Assessment of Originality}
Human assessment methods require at least two raters, with experience in the creative domain \cite{amabile_social_1982}, who independently assess the originality of ideas on a Likert scale \cite{silvia_assessing_2008, reiter-palmon_scoring_2019}. Raters are trained to score originality on a scale ranging from 1 (not at all creative) to 5 (highly creative).  As recommended by one of the most accepted scoring methods \cite{silvia_assessing_2008}, raters take into account the uncommonness (how infrequent an idea is within the sample of responses), remoteness (how conceptually distant the idea is from what is commonly thought), and cleverness (how smart, funny, insightful the idea is) of an idea \cite{agnoli_dopamine_2023, silvia_assessing_2008}. The interrater reliability (ICC) is then calculated on the total number of responses produced by participants. In case of important discrepancies between scores, the raters review their response and assign a score by consensus \cite{silvia_assessing_2008, agnoli_dopamine_2023}.

Although human-based methods are the core of creativity research \cite{reiter-palmon_scoring_2019, silvia_assessing_2008}, they present several notable limitations. First, given the substantial resources and effort required for raters' training and compensation, human scoring has become highly costly and extremely time-consuming \cite{saretzki_scoring_2025, distefano_automatic_2025, patterson_cap_2025}. This represents a significant obstacle for many researchers who do not have access to the necessary human resources \cite{zielinska_lost_2023, patterson_audra_2023, patterson_cap_2025}. Second, the assessment process may be influenced by rater fatigue, which can affect the reliability of their ratings \cite{forthmann_missing_2017}. Third, originality scores are affected by subjectivity and individual rater characteristics \cite{patterson_audra_2023, saretzki_scoring_2025}.

\subsection{Automatic Assessment of Originality}
To overcome these challenges, researchers have explored whether automated methods can replicate human assessment of idea ~originality, potentially offering a more objective and efficient solution \cite{organisciak_ocsai, beaty_semantic_2022, dumas2021measuring,beaty_automating_2021}. They can be divided into unsupervised and supervised methods.

\subsubsection{Unsupervised Methods}
Chronologically, one of the first attempts to automate originality scoring in divergent thinking tasks relied on unsupervised machine learning methods. They do not require human involvement and learning from previous examples \cite{beaty_semantic_2022, luchini_multilingual_2025, beaty_automating_2021}. In this regard, the most widely used approach is the semantic distance metric \cite{beaty_automating_2021, beaty_semantic_2022, dumas2021measuring}. It reflects how far apart two concepts are in the semantic space. 
Assessing creativity using this approach reflects the associative theory of creativity, which states that highly original ideas require the combination of semantically distant concepts (i.e., higher semantic distance values) \cite{beaty_associative_2023}. 

Semantic distance is quantified by $1- cosine$ of the angle between pairs of vectors in semantic space \cite{beaty_automating_2021, beaty_semantic_2022, luchini_automated_2025}. Considering the example provided by \cite{patterson2023multilingual}, \emph{coffee} and \emph{drink} have a low semantic distance ($.46$), whereas \emph{coffee} and \emph{write} have a higher semantic  distance ($.93$) \cite{patterson2023multilingual}. Semantic distance is considered a proxy of divergent thinking \cite{olson_naming_2021}.  
In the case of the AUT responses, semantic distance\footnote{Semantic distance can be computed through the publicly available platforms SemDis (https://cap.ist.psu.edu/semdis) and Open Creativity Scoring (https://openscoring.du.edu/scoring)} is computed as $1-$ the cosine angle between the target object (e.g, ``fork'' in the AUT) and the response (e.g., bottle opener)  \cite{beaty_semantic_2022, luchini_multilingual_2025, dumas2021measuring, beaty_automating_2021}. For instance\footnote{The following example is taken from https://openscoring.du.edu/scoring}, the angle between ``shovel'' and ``fling tennis balls for a dog to chase'' is $68$; the cosine of $68$ is $.37$, resulting in a semantic distance of $1 -.37 = .63$. Recent studies have shown that semantic distance predicts human originality scores in the AUT \cite{beaty_semantic_2022, dumas2021measuring, beaty_automating_2021}, with stronger correlations at the participant level but modest correlations at the response level \cite{organisciak_ocsai, luchini_automated_2025, saretzki_scoring_2025}.

\subsubsection{Supervised Methods}
Recently, LLMs have been proposed as an alternative approach for scoring idea originality in divergent thinking tasks. Trained LLMs have shown comparable performances to those of human raters in a variety of creative tasks, including the AUT \cite{organisciak_ocsai, zielinska_lost_2023}, metaphor generation \cite{distefano_automatic_2025}, and creative problem-solving tasks \cite{luchini_automated_2025}. Unlike semantic distance, LLM-based scoring represents a form of supervised learning. In this sense, LLMs are trained on previous input-output examples to predict output for new and previously unseen input. One of these models is OCSAI\footnote{https://openscoring.du.edu/scoringllm} (Open creativity scoring with artificial intelligence). 
The authors fine-tuned GPT-3 on 27,217 AUT responses and corresponding human ratings across nine datasets \cite{organisciak_ocsai}. Fine-tuning refers to the alignment of the model to a particular task, and in this case, fine-tuning allows the model to closely align with human ratings and strongly predict originality scores \cite{saretzki_scoring_2025,organisciak_ocsai, distefano_automatic_2025,luchini_automated_2025, patterson2023multilingual}. Recent studies found that OCSAI achieved high correlations with human raters (up to $r = .813$), outperforming semantic distance ($r = .20$) \cite{organisciak_ocsai, zielinska_lost_2023}. 

In August 2025, Patterson et al. \cite{patterson_cap_2025} introduced CAP (creativity assessment platform), which aims to provide an easy-to-use, flexible, and accessible platform. Creativity metrics include SemDis \cite{beaty_automating_2021} and five LLMs fine-tuned on previously collected data, each associated with a specific creativity measure (e.g., Short Stories, Scientific Creative Thinking, Design Problems). One of these models is CLAUS\footnote{https://cap.ist.psu.edu/claus} (Cross-Lingual Alternate Uses Scoring model), for scoring AUT responses. It employs a fine-tuned version of the XLM-RoBERTa language model to predict human creativity ratings for AUT responses and has been found to outperform SemDis  \cite{patterson_cap_2025}. CLAUS was trained on databases comprising 136,621 AUT responses and the corresponding human ratings in 12 languages~\cite{patterson_cap_2025}. 

\subsection{Comparative Studies of Originality Scoring}
Altogether, the studies reported above suggest that supervised methods are substantially faster than human raters while offering a comparable performance in originality scoring \cite{saretzki_scoring_2025, organisciak_ocsai, patterson_cap_2025}. However, little is known about the differences between OCSAI and CLAUS in scoring responses from divergent thinking tasks. In this regard, Saretzki et al. \cite{saretzki_scoring_2025} compared the performance of OCSAI, CLAUS, and GPT-4 in scoring AUT responses produced by \emph{human participants}. They found that models predictions outperformed those obtained with semantic distance, and were positively correlated with human ratings, with stronger relationships for OCSAI, followed by GPT-4 and CLAUS.  

Tang et al. \cite{Tang_2025} found a significant discrepancy between humans and OCSAI assessment. Unlike human raters, OCSAI overestimated the originality of divergent thinking task responses when they were generated by a human with the help of ChatGPT (Human-ChatGPT condition) compared to when they were produced by a couple of human participants (Human-Human condition). These findings suggest that OCSAI might prefer machine-generated responses, rating them as more original. To investigate the reasons behind this bias, the authors controlled for the elaboration of the participants' responses, operationalised as the number of characters. They hypothesised that OCSAI preferred responses of Human-ChatGPT condition because they were longer and more elaborated \cite{Tang_2025}. Controlling for the elaboration, they found that the advantage of the Human-ChatGPT condition over the Human-Human condition was significantly reduced. This finding was explained by hypothesising an elaboration bias in OCSAI, which overestimates originality scores of longer and more detailed responses.

To address the elaboration bias, Arora et al.~\cite{arora_generative_2025} proposed restricting the automatic assessment of OCSAI only to the core idea of machine-generated AUT responses. This strategy allowed them to analyse responses that were structurally similar to the ones produced by human participants. For instance, only the core idea ``Portable Zen Garden'' was extracted from the response: 

\begin{quote}
Portable Zen Garden, fill a shoe with fine sand, add a few tiny pebbles and a rake made from toothpicks. Instant relaxation for the foot-weary soul on the~go ~\cite{arora_generative_2025}.
\end{quote}

However, their findings contrasted with those reported by Tang et al. \cite{Tang_2025}. Indeed, they showed that OCSAI, despite controlling for elaboration, still rated the AUT responses generated by machines (ChatGPT-4o, DeepSeek-V3, and Gemini 2.0) as more original than those of human participants \cite{arora_generative_2025}. 

Overall, research on the validity and reliability of automated creativity assessment methods is still in its infancy. Some studies have examined the alignment between automated scores and human ratings in assessing the originality of human responses \cite{organisciak_ocsai, patterson_cap_2025, saretzki_scoring_2025, zielinska_lost_2023}. However, contrasting findings have emerged from recent studies investigating the performance of OCSAI \cite{Tang_2025, arora_generative_2025}. Thus, there is a lack of research systematically comparing different automatic systems (i.e., OCSAI, CLAUS, and ChatGPT) in their assessment of responses produced by \emph{both humans and LLMs}.

\section{Method} \label{sec_method}
This study aimed to extend previous research \cite{Tang_2025, arora_generative_2025} by investigating how OCSAI, CLAUS, and ChatGPT-4o align with human raters in scoring the originality of AUT ideas produced by ChatGPT-4o and human authors. 

\subsection{Database}
Human and machine responses, as well as the corresponding human rater originality scores, were obtained from the database of Domanti et al. \cite{Domanti_CHI2026}. The human sample comprised  81 psychology students (mean age = 20.3 years, SD = 1.71, men = 12, women = 69). Machine responses were collected on the 6th and 7th of March 2025 by using ChatGPT-4o chat interface\footnote{https://chat.openai.com}  with the ``Plus'' subscription \cite{Domanti_CHI2026}. This interface did not display or allow modification of hyperparameter settings. All data was collected in Italian.

Following psychological literature \cite{GUILFORD}, humans were asked to complete two AUT by generating as many original alternative uses as possible for ``shoe'' and ``fork'', with a time limit of 3 minutes for each target. 
The responses were recorded using an audio recorder and manually transcribed by one of the experimenters. The experimenter carefully listened to all recordings to ensure transcription accuracy and wrote them as expressed by the participants. The procedure was repeated by creating a new separate chat for each of the 81 ChatGPT-4o participants \cite{Domanti_CHI2026}. Specifically, ChatGPT-4o was prompted with the same instruction provided to human participants. Humans and ChatGPT-4o generated 4,813 valid AUT responses, of which 3,548 were generated by ChatGPT-4o and 1,265 by humans. ChatGPT-4o outperformed humans in fluency, producing an average of 43.8 ideas, compared to 15.6 for humans. 

In Domanti et al. \cite{Domanti_CHI2026}, after all the ideas were generated, all human and machine responses were merged and anonymised for assessment. Originality was scored by two expert coders, blind to the experimental goals and the origin of the data \cite{silvia_assessing_2008}. Accordingly, following the scoring procedures described in Silvia et al. \cite{silvia_assessing_2008}, both human raters underwent intensive training in which they were instructed to assess the originality of each response considering its \emph{uncommonness}, \emph{remoteness}, and \emph{cleverness} in a range of 1 (not all creative) to 5 (highly creative) \cite{Domanti_CHI2026}. Specifically, they were highly trained in the assessment of originality in divergent thinking tests until they reached a high level of consistency and agreement with expert raters, thus allowing them to be considered experts \cite{amabile_social_1982, silvia_assessing_2008}. In Domanti et al. \cite{Domanti_CHI2026}, the interrater reliability between the two raters was high, ICC = .93. 

Finally, following previous literature \cite[e.g.,][]{luchini_convergent_2023, Li03072021}, the human sample was divided into Higher Creative Humans (HCH) and Lower Creative Humans (LCH) via a median split based on the mean originality scores of each human participant. For example, an alternative use for a fork proposed by an HCH participant was to use it as a bridge in an anthill, with ants going from one side to the other. Conversely, an example provided by an LCH participant was using a fork to paint. The median split process is recommended to provide a more meaningful comparison of creativity \cite{Domanti_CHI2026}.  
\subsection{Automatic Assessment}\label{sec_assessment}
To collect automatic assessment data for the current study, we used OCSAI (version 1.6), CLAUS and ChatGPT-4o. Their ratings were collected on the 18th of December 2025. We also included ChatGPT-4o to investigate the capacity of a model to approximate human ratings without relying on any fine-tuning on the AUT \cite{saretzki_scoring_2025}. OCSAI scored  originality in a range of 1 to 5 \cite{organisciak_ocsai}. CLAUS scored the responses by providing a number from 0 to 1. It applies the following prompt in the background, where X is the target object and Y is the answer: \emph{Identify how surprising, creative, unexpected, or interesting the following alternative use for the object X is: Y} \cite{patterson_cap_2025, saretzki_scoring_2025}. 

ChatGPT-4o rating scores were collected through the API. We provided exactly the same instructions as for human raters. Consistent with \cite{saretzki_scoring_2025}, we prompted ChatGPT-4o to provide a single numerical value between 10 and 50, which allows a finer differentiation of the ratings compared to a 5-point Likert scale. ChatGPT-4o key model parameters, in particular temperature and top\_p, were controlled and kept at their default values of 1.0 and 1.0, respectively. CLAUS and ChatGPT-4o ratings were then rescaled to a 1-5 range to match the range of human raters and OCSAI and ensure their comparability \cite{saretzki_scoring_2025}. 
Finally, we repeated the median split procedure separately for each mean originality distribution obtained from OCSAI, CLAUS, and ChatGPT-4o. 
Supplementary Materials provide all responses generated by human and ChatGPT-4o, along with the corresponding originality scores assigned by the different raters. 
\section{Results}\label{sec_results}
We first analysed the originality of the ideas by considering the initial responses provided by humans and ChatGPT-4o (Section \ref{results_full_idea}). By initial response, we refer to the complete response provided by the participants at the AUT. Then, we controlled for the elaboration of the responses \cite{arora_generative_2025}, restricting the assessment only to the core idea (Section \ref{results_core_idea}). To differentiate between ChatGPT-4o and humans as authors and raters, in this section, we denote \GPTA and \HUMA as authors, and \GPTR and \HUMR as raters. 
\begin{table}[b]
\caption{Mean and Standard Deviation of initial response originality scores by rater and author}
\Description{Table 1 reports the means and standard deviations of initial response originality scores by author (ChatGPT-4o, lower creative humans, and higher creative humans) across four raters (Humans, OCSAI, CLAUS, and ChatGPT-4o)}
\label{tab:descriptive_full}
\centering
\begin{tabular}{lcccc}
\toprule
\textbf{Author} & \textbf{\HUMR} & \textbf{OCSAI} & \textbf{CLAUS} & \textbf{\GPTR} \\
\midrule
\GPTA & 1.99 (0.23) & 2.87 (0.55) & 3.31 (0.26) & 3.49 (0.69) \\
LCH   & 1.96 (0.23) & 2.62 (0.57) & 2.96 (0.27) & 2.37 (1.09) \\
HCH   & 2.11 (0.38) & 2.70 (0.62) & 3.09 (0.33) & 3.07 (1.10) \\
\bottomrule
\multicolumn{4}{l}{\footnotesize\emph{Note.} Values are reported as Mean (Standard Deviation).} \\
\end{tabular}
\end{table}
\subsection {Initial Response}\label{results_full_idea}
Correlations of the mean originality scores between the automatic systems were moderate to strong, ranging from .61 (OCSAI – \GPTR and OCSAI - CLAUS) to .77 (\GPTR – CLAUS). When considering only machine-generated responses, agreement was lower for OCSAI – \GPTR ($r = .38$) and \GPTR – CLAUS ($r = .49$), while OCSAI - CLAUS remained moderate ($r = .62$). Overall, human ratings showed a weak but significant positive association with OCSAI ($r = .28$), which became robust ($r = .59$) when considering only human responses.
\subsubsection{Frequencies}
Figure \ref{fig:full_density} illustrates the kernel density estimates of originality scores of ideas produced by \HUMA and \GPTA, as assessed from \HUMR, OCSAI, CLAUS and \GPTR. It shows how originality scores are distributed differently across the scale depending on the rater.  

\HUMR were concentrated around a single scale point, resulting in a highly right-skewed distribution, with 94\% of all the scores rated 2.0. Specifically, for \GPTA,  98\% of the scores fell between 1.0 and 2.0, whereas the percentage for \HUMA was 92\%. 6 responses generated by \GPTA and 13 generated \HUMA fell between 4.0 and 5.0. OCSAI distribution was characterised by a central tendency \cite{saretzki_scoring_2025}, with 60\% of all observations concentrated at the value 3.0. Considering only \GPTA productions, 90\% of scores fell between 2.0 and 3.0. \HUMA exhibited a lower peak at the value 3.0 with a percentage of 54\%.
Consistent with \cite{saretzki_scoring_2025}, CLAUS scores exhibited a more symmetric and narrower distribution, approximating a normal distribution. For \GPTA, 85\% of scores fell between 3.00 and 3.76, while for \HUMA 85\% of scores were contained within the interval from  2.72 to 3.52. 
Finally, when \GPTR assessed the responses, originality scores were widely distributed on the scale, with no value exceeding 20\% of the observations. In this case, 63\% of \GPTA scores fell between 3.0 and 4.0, whereas for \HUMA it was substantially lower (37\%).
\begin{table*}[b]
  \caption{Initial response originality results by rater}
  \Description{Table 2 reports initial response originality comparison results by rater (Humans, OCSAI, CLAUS, and ChatGPT-4o). For each rater, three pairwise comparisons are shown (ChatGPT-4o vs lower creative humans, ChatGPT-4o vs higher creative humans, and higher creative humans vs lower creative humans), with estimates B, standard errors, 95\% confidence intervals, t-statistics, and p-values.}
  \label{tab:full_lmm}
  \begin{tabular}{cllllll}
    \toprule
    \textbf{Rater} & \textbf{Comparison} & \textbf{B} & \textbf{SE} & \textbf{95\%~CI} & \textbf{t-statistic} & \textbf{p-value}\\
    \midrule
    \multirow{3}{*}{\shortstack[l]{\HUMR}} 
      & \GPTA\ $>$ \ LCH & $0.03$ & $0.013$ & $[0.00, 0.05]$ & $t_{(359.46)} = 1.98$ & $p = .048$ \\
      & \GPTA\ $<$ \ HCH &  $- 0.12$ & $0.011$ & $[-0.14, -0.10]$ & $t_{(193.59)} = -11.35$ & $p < .001$ \\
      & HCH\ $>$ \ LCH   & $0.15$ & $0.016$ & $[0.12, 0.18]$ & $t_{(391.45)} = 9.51$ & $p < .001$ \\
\midrule
\multirow{3}{*}{\shortstack[l]{OCSAI}} 
      & \GPTA\ $>$ \ LCH & $0.25$ & $0.025$ & $[0.20, 0.30]$ & $t_{(4809.99)} = 9.69$ & $p < .001$ \\
      & \GPTA\ $>$ \ HCH &  $0.17$ & $0.023$ & $[0.12, 0.21]$ & $t_{(4809.31)} = 7.08$ & $p < .001$ \\
      & HCH\ $>$ \ LCH   & $0.08$ & $0.032$ & $[0.02, 0.14]$ & $t_{(4809.36)} = 2.55$ & $p = .011$ \\
      \midrule
\multirow{3}{*}{\shortstack[l]{CLAUS}} 
      & \GPTA\ $>$ \ LCH &  $0.35$ & $0.012$ & $[0.33, 0.38]$ & $t_{(313.09)} = 28.50$ & $p < .001$ \\
      & \GPTA\ $>$ \ HCH &  $0.22$ & $0.012$ & $[0.19, 0.24]$ & $t_{(317.23)} = 17.83$ & $p < .001$ \\
      & HCH\ $>$ \ LCH & $0.14$ & $0.016$ & $[0.10, 0.17]$ & $t_{(435.84)} = 8.54$ & $p < .001$ \\
    \midrule
\multirow{3}{*}{\shortstack[l]{\GPTR}} 
      & \GPTA\ $>$ \ LCH &  $1.11$ & $0.041$ & $[1.03, 1.20]$ & $t_{(245.89)} = 27.29$ & $p < .001$ \\
      & \GPTA\ $>$ \ HCH &  $0.42$ & $0.039$ & $[0.34, 0.49]$ & $t_{(215.12)} = 10.68$ & $p < .001$\\
      & HCH\ $>$ \ LCH   & $0.70$ & $0.051$ & $[0.60, 0.80]$ & $t_{(311.80)} = 13.69$ & $p < .001$ \\
    \bottomrule
  \end{tabular}
\end{table*}
\begin{figure*}[t]
    \centering

    \begin{subfigure}{0.48\textwidth}
        \centering
        \includegraphics[width=\linewidth]{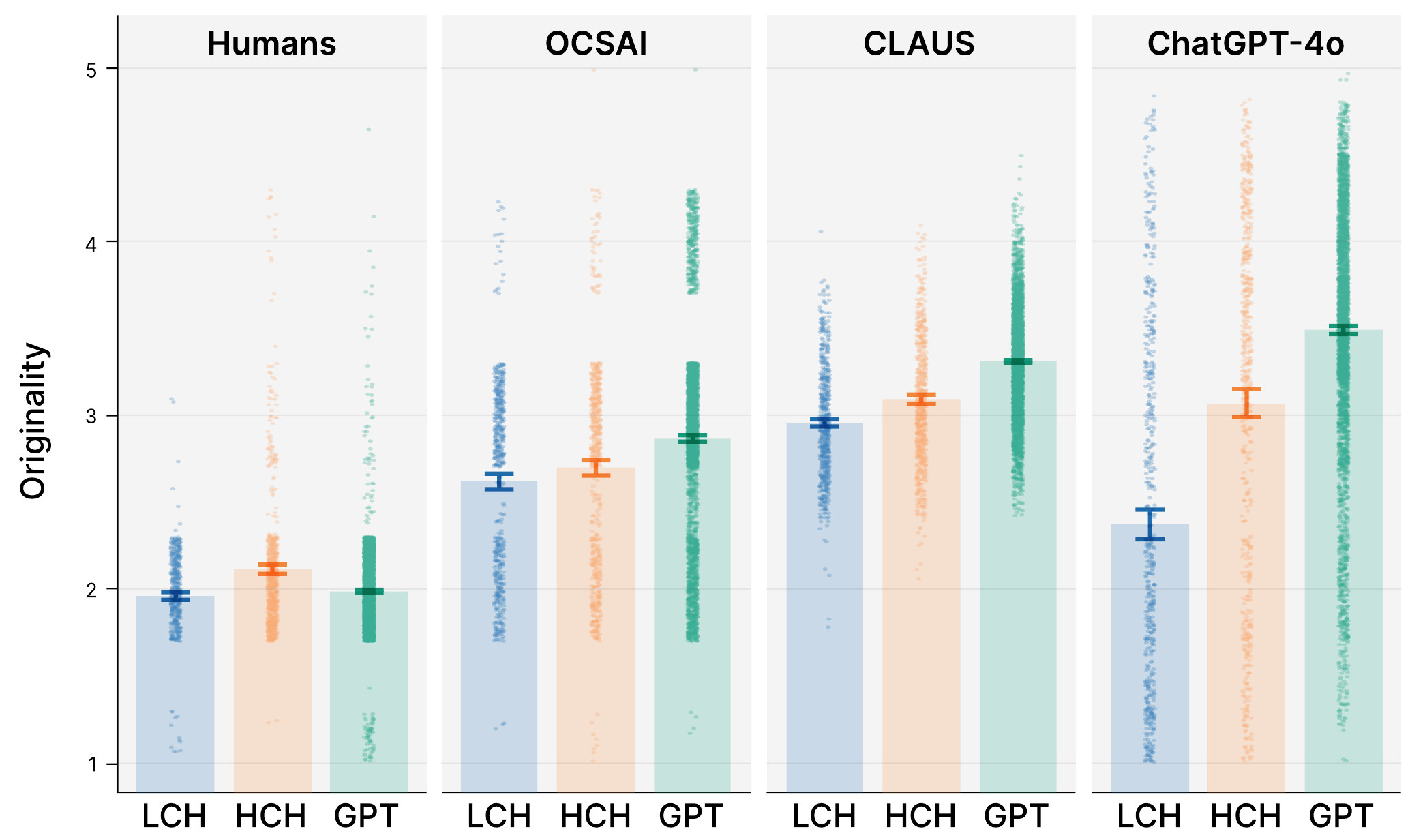}
        \caption{}
        \label{fig:PLOT_FULL}
    \end{subfigure}
    \hfill
    \begin{subfigure}{0.48\textwidth}
        \centering
        \includegraphics[width=\linewidth]{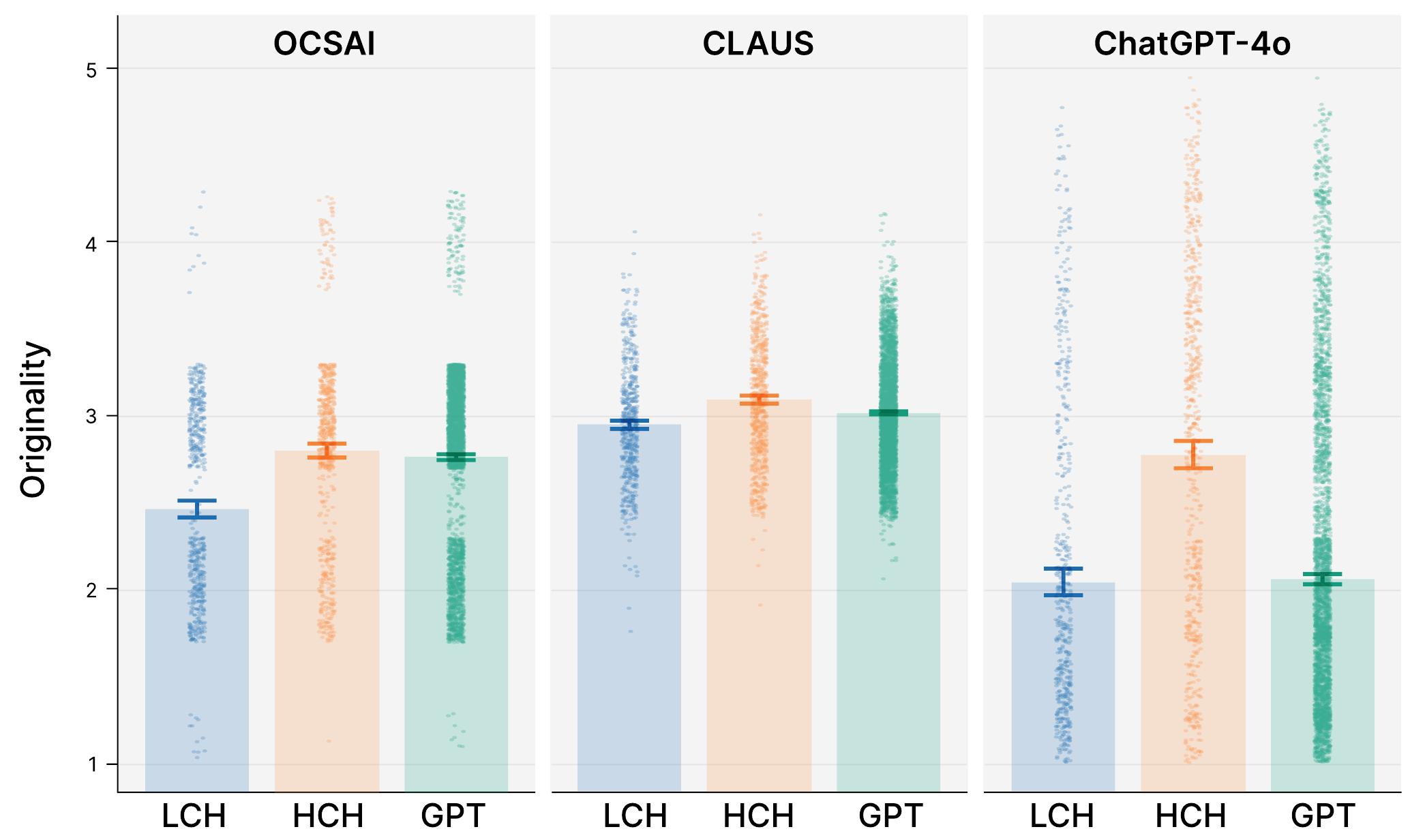}
        \caption{}
        \label{fig:PLOT_SHORT}
    \end{subfigure}

    \caption{(a): Initial response originality scores across authors and raters (Humans, OCSAI, CLAUS, ChatGPT-4o). (b): Core idea originality scores across authors and raters (OCSAI, CLAUS, ChatGPT-4o)}
    
    \Description{Two side-by-side panels. The left panel (a)  shows initial response originality scores for three groups (LCH, HCH, and ChatGPT-4o) across four raters (Humans, OCSAI, CLAUS, and ChatGPT-4o), displayed in four panels, one per rater. The Y axis shows originality scores from 1 (low) to 5 (high). Within each panel, the three groups are plotted on the X axis (lower creative humans, higher creative humans, ChatGPT-4o). For each group, data are shown as dots and there are error bars indicating the group average and uncertainty. For human raters, higher creative humans are more original than ChatGPT-4o, which is more original than lower creative humans. For OCSAI, CLAUS, ChatGPT-4o raters, ChatGPT-4o is more original than higher creative humans, who are more original than lower creative humans. The right panel (b) shows core response originality scores for three groups (Lower creative humans, higher creative humans, and ChatGPT-4o) across three raters (OCSAI, CLAUS, and ChatGPT-4o), displayed in three panels, one per rater. The Y axis shows originality scores from 1 (low) to 5 (high). Within each panel, the three groups are plotted on the X axis (lower creative humans, higher creative humans, ChatGPT-4o). For each group, data are shown as dots, and there are error bars indicating the group average and uncertainty. For OCSAI and CLAUS, higher creative humans are more original than ChatGPT-4o, which is more original than lower creative humans. For ChatGPT-4o as a rater,  higher creative humans are more original than ChatGPT-4o, but there is no difference in originality between ChatGPT-4o and lower creative humans.}

    \label{fig:combined_originality}
\end{figure*}
\subsubsection{Originality} \label{originality_LMM}
We then calculated descriptive statistics of initial response originality (Table~ \ref{tab:descriptive_full}) and ran linear mixed models to investigate differences in originality among the three groups for each rater. The analyses were run with the lme4 package \cite{bates_fitting_2015} and lmerTest \cite{kuznetsova_lmertest_2017} in R. For these analyses, we considered the human sample split in LCH and HCH, as explained in the Method (Section \ref{sec_method}). The dependent variable consisted of all responses rated and the fixed effect was Participant (\GPTA, LCH, HCH), while intercepts for participants’ ID and object (shoe and fork) served as random effects. 
Figure \ref{fig:PLOT_FULL} shows the AUT originality scores across authors (LCH, HCH, \GPTA) and raters (\HUMR, OCSAI, CLAUS, \GPTR). In Domanti et al. \cite{Domanti_CHI2026}, results showed that, for \HUMR, \GPTA originality scores were significantly higher than those of LCH, but lower than those of HCH. Furthermore, HCH originality scores were significantly higher than those of LCH. 

Regarding automatic assessment, our results showed a reverse pattern. They suggest a self-preference bias, whereby machine-generated responses were preferred over human responses. Indeed, OCSAI, CLAUS, and \GPTR agreed in recognising \GPTA as the most original author. For all of them, \GPTA originality scores were significantly higher than those of LCH and HCH. In accordance with \HUMR, they rated HCH as significantly more original than LCH. Importantly, for OCSAI, the random intercept for participants’ ID was removed due to a singular fit and zero estimated variance. Table \ref{tab:full_lmm} provides a summary of the results of the linear mixed models.
\begin{table}[h]
\caption{Mean and Standard Deviation of core idea originality scores by rater and author}
\Description{Table 3 reports the means and standard deviations of core response originality scores by author (ChatGPT-4o, lower creative humans, and higher creative humans) across three raters (OCSAI, CLAUS, and ChatGPT-4o)}
\label{tab:descriptiveSHORT}
\centering
\begin{tabular}{lccc}
\toprule
\textbf{Author} & \textbf{OCSAI} & \textbf{CLAUS} & \textbf{\GPTR} \\
\midrule
\GPTA & 2.77 (0.49) & 3.02 (0.26) & 2.06 (0.91) \\
LCH   & 2.46 (0.59) & 2.95 (0.28) & 2.04 (0.98) \\
HCH   & 2.80 (0.56) & 3.10 (0.32) & 2.78 (1.08) \\
\bottomrule
\multicolumn{4}{l}{\footnotesize\emph{Note.} Values are reported as Mean (Standard Deviation).} \\
\end{tabular}
\end{table}
\subsection {Core Idea} \label{results_core_idea}
To investigate the LLM self-preference bias observed in OCSAI, CLAUS, and \GPTR, we controlled for the elaboration of responses. Accordingly, based on \cite{arora_generative_2025}, we extracted only the core idea from each initial response. Importantly, to ensure a comparable assessment of machine and human responses, some \HUMA responses starting with ``as'' (e.g., as small bag) were modified by keeping only the core idea (e.g., small bag). 
\subsubsection {Frequencies}
\begin{figure*}[t]
    \centering
    \includegraphics[width=\textwidth]{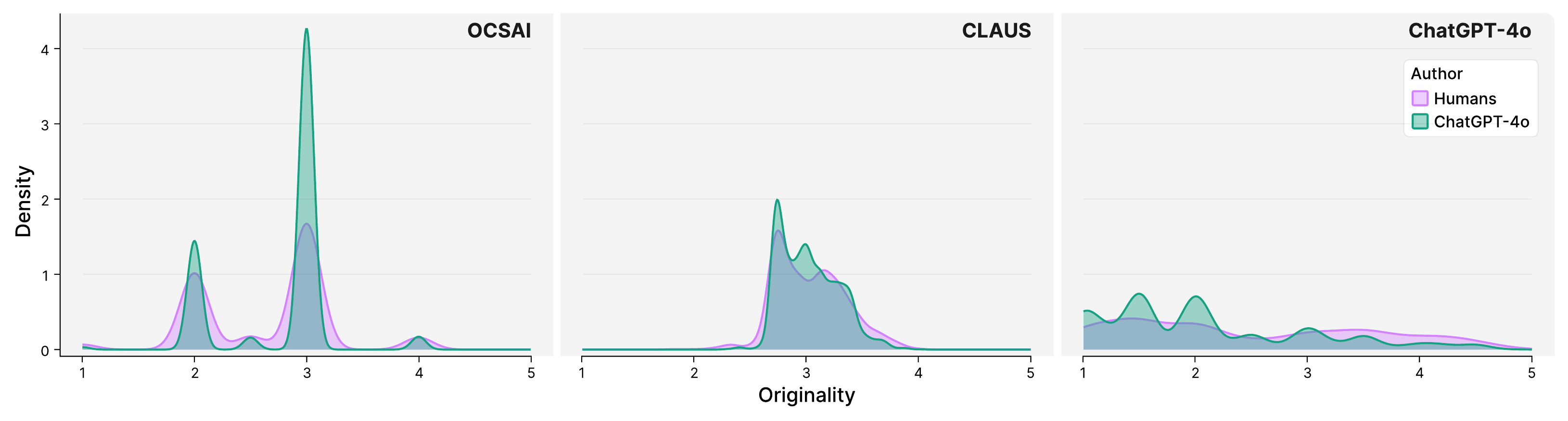}
    \Description{Figure 3 shows kernel density estimates of Alternate Uses Task originality scores of Human participants and ChatGPT-4o, shown separately for three raters: OCSAI, CLAUS, and ChatGPT-4o. Each panel displays two density curves (Humans and ChatGPT-4o). The X-axis shows the originality scale and it ranges from 1 to 5. The Y-axis represents the density. It ranges from 0 to 4. For OCSAI, the two curves have peaks around 2.0 and 3.0. For CLAUS, the two curves are narrower and largely overlapping around the mid-range. For ChatGPT-4o as a rater, the two curves are widely distributed on the scale.}
    \caption{Kernel density estimates of core idea originality scores across raters (OCSAI, CLAUS, ChatGPT-4o)}
    \label{fig:p_density_short}
\end{figure*}

Figure \ref{fig:p_density_short} illustrates the kernel density estimates of originality scores of \HUMA and \GPTA ideas, as obtained from OCSAI, CLAUS and \GPTR. Looking at the frequencies, 96\% of OCSAI scores fell between 2.0 and 3.0. In particular, 70\% of \GPTA scores and 54\% of \HUMA scores were rated 3.0. CLAUS showed a narrower distribution \cite{saretzki_scoring_2025}. In general, \GPTA and \HUMA largely overlapped and reached their peak at 2.72. Nevertheless, 11\% of \HUMA productions received scores above 3.4, compared to only 6\% of \GPTA productions. Finally, for \GPTR, the originality scores were again more widely distributed on the scale, with no value exceeding 22\% of the observations. Overall, 81\% of scores fell between 1.0 and 3.0. For \GPTA, 47\% of scores fell between 1.5 and 2.0, while for \HUMA it was lower (35\%).
\subsubsection {Originality}
Descriptive statistics of core idea originality are reported in Table \ref{tab:descriptiveSHORT}. We ran linear mixed models to investigate differences in originality among the three groups for each rater, following the same procedure described in Subsection \ref{originality_LMM}. Importantly, due to a singular fit and zero estimated variance, the random intercept for participants’ ID was removed for OCSAI, and the random intercepts for participants’ ID and object were removed for CLAUS. 
AUT originality scores across authors (LCH, HCH, \GPTA) and raters (OCSAI, CLAUS, \GPTR) are shown in Figure~\ref{fig:PLOT_SHORT}.  The results showed a change in the ratings of OCSAI, CLAUS, and \GPTR, with HCH rated as significantly more original than \GPTA (this effect was only marginally significant for OCSAI). Moreover, whereas OCSAI and CLAUS rated \GPTA as more original than LCH, this difference did not emerge when ChatGPT-4o was the rater. Table \ref{tab:short_lmm} provides a summary of the results of the linear mixed models.
\begin{table*}[b!]
  \caption{Core idea originality results by rater}
  \Description{Table 4 reports core response originality comparison results by rater (OCSAI, CLAUS, and ChatGPT-4o). For each rater, three pairwise comparisons are shown (ChatGPT-4o vs lower creative humans, ChatGPT-4o vs higher creative humans, and higher creative humans vs lower creative humans), with estimates B, standard errors, 95\% confidence intervals, t-statistics, and p-values.}
  \label{tab:short_lmm}
  \begin{tabular}{cllllll}
    \toprule
    \textbf{Rater} & \textbf{Comparison} & \textbf{B} & \textbf{SE} & \textbf{95\%~CI} & \textbf{t-statistic} & \textbf{p-value}\\
    \midrule
\multirow{3}{*}{\shortstack[l]{OCSAI}} 
      & \GPTA $>$ LCH &  $0.30$ & $0.023$ & $[0.25, 0.34]$ & $t_{(4809.04)} = 12.95$ & $p < .001$ \\
      & \GPTA $<$* HCH & $-0.04$ & $0.021$ & $[-0.08, 0.00]$ & $t_{(4809.02)} = -1.95$ & $p = .051$\\
      & HCH $>$ LCH   & $0.34$ & $0.029$ & $[0.28, 0.39]$ & $t_{(4809)} = 11.78$ &  $p < .001$ \\
      \midrule
\multirow{3}{*}{\shortstack[l]{CLAUS}} 
      & \GPTA\ $>$ \ LCH &  $0.07$ & $0.012$ & $[0.05, 0.10]$ & $t_{(4810)} = 5.96$ & $p < .001$ \\
      & \GPTA\ $<$ \ HCH &   $-0.07$ & $0.011$ & $[-0.10, -0.05]$ & $t_{(4810)} = -6.72$ & $p < .001$ \\
      & HCH\ $>$ \ LCH   & $0.15$ & $0.015$ & $[0.12, 0.18]$ & $t_{(4810)} = 9.68$, & $p < .001$ \\
    \midrule
\multirow{3}{*}{\shortstack[l]{\GPTR}} 
      & \GPTA\ $=$ \ LCH &  $0.02$ & $0.042$ & $[-0.06, 0.10]$ & $t_{(276.46)} = 0.44$ & $p = .66$ \\
      & \GPTA\ $<$ \ HCH &   $-0.73$ & $0.040$ & $[-0.80, -0.65]$ & $t_{(224.93)} = -18.11$ & $p < .001$\\
      & HCH\ $>$ \ LCH   & $0.74$ & $0.053$ & $[0.64, 0.85]$ & $t_{(352.65)} = 13.92$ & $p < .001$ \\
    \bottomrule
     \multicolumn{2}{l}{* marginally significant}
  \end{tabular}
\end{table*}
\section{Discussion}\label{sec_discussion}
LLMs are increasingly used as authors \cite[e.g.,][]{stevenson_2022putting, HAASE_2023, koivisto_best_2023} and raters \cite[e.g.,][]{organisciak_ocsai, patterson_cap_2025, saretzki_scoring_2025} of responses in creative tasks. The study reported in this paper focused on creativity assessment:  the use of LLMs to score originality. Dealing with large databases fuelled by automatic generation, this possibility may represent a promising and efficient alternative to subjective and costly human-based assessment. However, important concerns remain about validity, as these systems have been shown to be influenced by biases in scoring procedures \cite{Tang_2025, self_llm_2024}. Research is still in its infancy and only a few studies have investigated the alignment between LLMs and human scores of idea originality on the AUT \cite[e.g.,][]{saretzki_scoring_2025, zielinska_lost_2023}. Furthermore, uncertainty exists about the performance of the LLM automatic systems in scoring responses generated by LLMs. 

Therefore, this study investigated how OCSAI, CLAUS, and ChatGPT-4o align with human raters in scoring the originality of AUT responses provided by humans and ChatGPT-4o. Table \ref{tab:summary} summarises the results.
\subsection{Comparison of Raters}
Focusing on the distributions of the originality scores in the initial responses across raters, our results revealed significant differences in how the values were distributed on the originality scale as a function of the rater  (Figure \ref{fig:full_density}). The distribution obtained from human raters shows that only a very small number of unique ideas were recognised as highly creative, while the majority fell below the mean value \cite{saretzki_scoring_2025}. In contrast, LLM systems tended to assign higher originality scores than human raters (Figure \ref{fig:PLOT_FULL}). Consistent with previous work \cite{saretzki_scoring_2025}, we found that OCSAI exhibited a central tendency around the value of 3.0, whereas CLAUS showed a more symmetric and narrower distribution. ChatGPT-4o, by contrast, was not characterised by a central tendency, but instead distributed the scores along the entire originality scale. 

Going deeper into our results,  human raters assessed ChatGPT-4o as less original than higher creative humans, but more original than lower creative ones. This finding is consistent with recent literature suggesting that a percentage of the best humans outperform LLMs in divergent thinking tasks \cite{koivisto_best_2023, HAASE_2023, grassini_artificial_2025}. 
However, in line with Tang et al. \cite{Tang_2025}, our results confirmed the presence of a self-preference bias in LLMs.  OCSAI, CLAUS, and ChatGPT-4o assessed ChatGPT-4o responses as more original than those generated by both human groups \cite{hubert_current_2024, GUZIK2023100065}. Thus, they preferred machine-generated responses, assigning them higher originality scores \cite{self_llm_2024}. Among them, OCSAI and CLAUS showed similar patterns, with the latter assigning higher originality scores than the former (Figure \ref{fig:PLOT_FULL}). ChatGPT-4o exhibited a wider range of scores than OCSAI and CLAUS, despite not being fine-tuned on AUT responses.

Inspired by \cite{arora_generative_2025}, we controlled for the response elaboration by considering only the core idea of each initial response to investigate the self-preference bias observed in LLM raters \cite{Tang_2025, self_llm_2024}.  Our results extend previous findings \cite{Tang_2025, arora_generative_2025}, showing that OCSAI, CLAUS, and ChatGPT-4o assessed ChatGPT-4o core ideas as less original than those generated by higher creative individuals (even if this effect was only marginally significant for OCSAI). Thus, they recognised higher creative humans as the most original group. Importantly, ChatGPT-4o was the most affected rater by the elaboration control. Indeed, the mean originality of ChatGPT-4o responses decreased by 1.43 points, going from 3.49 for the initial response to 2.06 for the core idea. Additionally, while OCSAI and CLAUS rated machine core ideas as more original than those of lower creative humans, ChatGPT-4o did not score them as more original than those of lower creative humans (Table \ref{tab:summary}). 
\begin{table}[t]
  \caption{Summary of results}
  \Description{Table 5 presents a summary of the results of the initial response and core idea by rater}
  \label{tab:summary}
  \begin{tabular}{p{1.3cm} p{3cm} p{3cm}}
    \toprule
    \textbf{Rater} & \textbf{Initial Response} & \textbf{Core Idea} \\
    \midrule
    \HUMR  & HCH $>$ \GPTA $>$ LCH & \multicolumn{1}{c}{/} \\
    OCSAI & \GPTA $>$ HCH $>$ LCH & HCH $>$* \GPTA $>$ LCH \\
    CLAUS & \GPTA $>$ HCH $>$ LCH & HCH $>$ \GPTA $>$ LCH \\
    \GPTR  & \GPTA $>$ HCH $>$ LCH & HCH $>$ \GPTA $=$ LCH \\
    \bottomrule
    \multicolumn{3}{l}{* marginally significant}
  \end{tabular}
\end{table}
\subsection{Implications for Theory and Method}
Overall, our results raise concerns about the appropriateness of using LLM systems to score the originality of responses generated by LLMs. They highlight the challenges associated with automating creativity scoring and provide evidence of potential elaboration and self-preference biases exhibited by these systems \cite{patterson_cap_2025, Tang_2025}. These findings have both theoretical and methodological implications for ~HCI. 

From a theoretical perspective, our results extend the literature on risk of homogenisation associated with the pervasive influence of AI on human creativity \cite{Anderson, doshi_generative_nodate, Kumar}. For example, we observed zero estimated variance for participants random intercept when OCSAI assessed both initial responses and core ideas. This result indicates a homogenisation effect at the participant level in the assessment of OCSAI. Moreover, our findings reveal challenges for researchers and users. Relying on LLM scoring systems that are biased towards their own productions can inflate originality scores, introducing error into the results and ``making the literature noisy'' ~\cite[p. 18]{Tang_2025}. From the point of view of the users,  this dynamic may foster Creative Mortification \cite{beghetto_creative_2014}, leading them to perceive their outcomes as less valuable than those generated by AI systems.

Given that idea generation is increasingly shaped by the ultra-rapid speed at which LLMs generate content, the assessment phase becomes essential to avoid negative consequences such as hallucinations and homogenisation \cite{beaty_chapter_2026, Anderson, doshi_generative_nodate, Kumar}. Our findings suggest that removing the human judgment from the originality assessment could be risky, resulting in inappropriate and biased scores of creative outcomes. Furthermore,  emotional depth, lived experience, and contextual dimension of creativity remain essential to comprehensively assess creativity, including its usefulness or appropriateness, beyond originality alone. These concerns represent a challenge that the entire field of research should address and provide support for the differentiation between artificial creativity and human creativity \cite{RUNCO_2023}. 

From a methodological point of view, our findings suggest that applying a split in the human sample is a useful strategy to obtain a finer-grained operationalisation of creativity \cite{Domanti_CHI2026}. Rather than treating humans as a single group, this approach brings the variability of human performance to the foreground. Human creativity is not an average skill, but requires a comparison that considers both higher and lower levels of performance \cite{Domanti_CHI2026, kenett_investigating_2014}. A limitation of prior work is that machines are often compared against a single and undifferentiated human group. This methodological choice can be misleading and hide the intrinsic differences in the human sample. Accordingly, when a machine outperforms the human sample in creative tasks, the conclusion is that machines are more original than humans \cite[e.g.,][]{arora_generative_2025, hubert_current_2024}. In contrast, splitting the human sample can reveal that higher creative humans still outperform the machine, which instead performs at approximately the average human level \cite{cropley_averagecreativity_2025}.

\subsection{Limitations and Future Directions}
The study presents some limitations. Firstly, we focused only on the AUT as a measure of divergent thinking. Although this task is a standard in the investigation of human and artificial creativity \cite[e.g.,][]{Bangerl_2025, Kumar, stevenson_2022putting, HAASE_2023, arora_generative_2025, koivisto_best_2023}, future studies should extend the investigation by comparing rater performances across different creative thinking tasks, including creative story writing, metaphor production, or non-verbal tasks. 

Secondly,  we used ChatGPT-4o to assess the originality of the responses alongside OCSAI and CLAUS (specialised systems). Building on our methodology, future research could examine the performance of other LLMs, both as participants and as raters. Moreover, we provided ChatGPT-4o with the same instructions given to human raters, emphasising the uncommonness, remoteness, and cleverness of an idea \cite{silvia_assessing_2008}. Furthermore, future work is needed to investigate how different prompting strategies affect the automatic scoring of originality. For instance, future studies might ask an LLM to score responses using the same prompt that CLAUS applies in its assessment (Section \ref{sec_assessment}), or manipulate model hyperparameter settings (e.g., temperature, top\_p) to explore their influence on the assessment.

Thirdly, we addressed the LLMs tendency to prefer their own ideas by adopting the strategy proposed by Arora et al. \cite{arora_generative_2025}. Future research should investigate whether and how human raters are influenced by the elaboration bias. It would be interesting to explore how the elaboration control proposed by Tang et al. \cite{Tang_2025} affects the performance of other LLMs and human raters. Finally, since our responses were in Italian, we prompted ChatGPT-4o accordingly; CLAUS supports this language. To ensure a reliable comparison, we used the multilingual version of OCSAI. Evidence on LLM multilingual performance in originality scoring is still limited \cite{saretzki_scoring_2025, zielinska_lost_2023}. Therefore, future research in this field should compare how the language of the responses influences the assessment performance of different LLMs.

\section{Conclusion}\label{sec_conclusion}
To conclude, this paper investigated the alignment of LLMs with humans in assessing the idea originality in the Alternate Uses Task. Going back to our research question, we provide evidence of a self-preference bias in LLMs. All automatic systems (OCSAI, CLAUS, and ChatGPT-4o) preferred machine-generated responses over human ideas, but this pattern reversed after controlling for the idea elaboration. Altogether, these results provide empirical support for the ``Shut down'' scenario envisioned by Vinchon et al. \cite{vinchon_manifesto_2023}. Indeed, by preferring artificial responses, automatic assessment might discourage users from starting a creative action. Finally, this paper contributes to the research challenges proposed in the decalogue for cyber-creativity \cite{Corazza_decalogue},  highlighting how automatic scoring can introduce biases into the assessment process.

\begin{acks}
We thank Fe Simeoni for help with the design of the figures. The second author is funded by the European Union- Next Generation EU, Mission 4 Component 1 CUP I52B23000570003. We acknowledge financial support by the European Union – NextGenerationEU, CUP J53D2300799 0001.
\end{acks}

\balance
\bibliographystyle{ACM-Reference-Format}
\bibliography{sample-base}
\end{document}